\documentclass{proc}
\usepackage{amsfonts}
\usepackage{amsmath}
\usepackage{amssymb}

\usepackage{xspace}
\usepackage{graphicx}
\usepackage{algorithm}
\usepackage{algorithmic}

\usepackage{float}

  {\endminipage\par\medskip}

\newcommand{\smallbsq}{\hfill{\tiny $\blacksquare$}} 

\def\kras{\mathit{krasnovia}}
\def\baja{\mathit{baja}}
\def\mojave{\mathit{mojave}}
\def\worm{\mathit{worm123}}
\def\mwsam{\mathit{mw123sam1}}
\def\mwsamT{\mathit{mw123sam2}}
\def\eng{\mathit{english}}

\def\Tr{\textsf{True}}

\def\f{{\theta}}
\def\r{{\omega}}
\def\p{{{\phi}}}
\def\d{{{\delta}}}

\def\evidOf{\mathit{evidOf}}
\def\isCap{\mathit{isCap}}
\def\motiv{\mathit{motiv}}
\def\tgt{\mathit{tgt}}
\def\condOp{\mathit{condOp}}
\def\hasMse{\mathit{hasMseInvest}}

\def\expCw{\mathit{expCw}}

\def\compiledLan{\mathit{compilLang}}
\def\mwInOp{\mathit{malwInOp}}
\def\natLang{\mathit{nativLang}}
\def\mwHint{\mathit{mwHint}}
\def\origIP{\mathit{origIP}}
\def\inLgConf{\mathit{inLgConf}}
\def\mwRel{\mathit{malwareRel}}
\def\cooper{\mathit{cooper}}
\def\infSys{\mathit{infGovSys}}
\def\numTtMse{\mathit{mseTT}}
\def\cybCapAge{\mathit{cybCapAge}}
\def\govCybCap{\mathit{govCybLab}}

\def\act{\textsf{A}}
\def\Cons{\textsf{\textbf{C}}}
\def\actors{\Cons_{act}}
\def\ops{\Cons_{ops}}
\def\Var{\textsf{\textbf{{V}}}}
\def\PredE{\textsf{\textbf{{P}}}_\textit{{EM}}}
\def\PredA{\textsf{\textbf{{P}}}_\textit{{AM}}}
\def\GndE{\textsf{\textbf{{G}}}_\textit{{EM}}}
\def\GndA{\textsf{\textbf{{G}}}_\textit{{AM}}}
\def\wldE{\mathcal{W}_\textit{{EM}}}
\def\wldA{\mathcal{W}_\textit{{AM}}}
\def\cala{\mathcal{A}}
\def\ICE{\textsf{\textbf{{IC}}}_\textit{{EM}}}
\def\ICA{\textsf{\textbf{{IC}}}_\textit{{AM}}}

\def\Pr{\textsf{\textit{Pr}}}

\def\PAM{\Pi_{\mathit{AM}}}
\def\PEM{\Pi_{\mathit{EM}}}
\def\kE{\PEM}
\def\cali{\mathcal{I}}
\def\fEM{formula_{EM}}
\def\fAM{formula_{AM}}
\def\war{\vdash_{\textsf{war}}}
\def\nwar{\not\vdash_{\textsf{war}}}
\def\must{nec}
\def\can{poss}

\def\calf{\mathcal{F}}
\def\calfs{\calf^*}
\def\<{\langle}
\def\>{\rangle}
\def\calt{\mathcal{T}}
\def\calts{\mathcal{T}^*}
\def\for{for}
\def\af{\mathit{af}}
\def\calo{\O}
\def\Pdist{\textbf{\textsf{P}}_{L,\Pr,\cali}}
\def\Pkb{\textbf{\textsf{P}}_{L,\cali}}

\def\Pkbco{\textbf{\textsf{P}}_{\condOp(\act,\calo),\cali'}}
\def\Pkbpco{\textbf{\textsf{P}}_{\condOp(\act',\calo),\cali'}}

\newtheorem{example}{Example}[section]
\newtheorem{definition}{Definition}[section]

\newcommand{\SRules}{\mbox{$\Omega$}}
\newcommand{\Facts}{\mbox{$\Theta$}}
\newcommand{\DRules}{\mbox{$\Delta$}}
\newcommand{\Presumps}{\mbox{$\Phi$}}
\newcommand{\pdelpprog}{(\Facts, \SRules, \Presumps,  \DRules)}
\newcommand{\PDeLP}{PreDeLP}

\newcommand{\DLP}{\mbox{{\sc DeLP}}}

\newcommand{\no}{\mbox{$\neg$}}
\newcommand{\PP}{\PAM}
\newcommand{\DD}{\mbox{$\DRules\cup\Presumps$}}

\newcommand{\Prog}{\mbox{$\PP$}}

\newcommand{\Arg}{\mbox{$\mathcal{A}$}}
\newcommand{\AL}{\ensuremath{\langle \Arg,L \rangle}}

\newcommand{\srule}[2]{\mbox{$#1\!\leftarrow#2$}}

\newcommand{\defleft}{\mbox{\bf--\hspace{-1pt}\raise.1185pt\hbox{$\prec$} }}
\newcommand{\defleftarrow}{{\raise1.5pt\hbox{\tiny\defleft}}}
\newcommand{\drule}[2]{\mbox{$#1 \;\defleftarrow #2$}}

\newcommand{\Tree}[1]{\mbox{${\mathcal T}({\small #1})$}}
\newcommand{\MTree}[1]{\mbox{${\mathcal T^*}({\small #1})$}}

\newcommand{\Barg}{\mbox{${\mathcal B}$}}

\newcommand{\Bq}{\mbox{$\langle \Barg,q \rangle $}}
\newcommand{\Ah}{\mbox{$\langle \Arg,h \rangle $}}

\newcommand{\Dnode}{\mbox{``\textsf{D}''}}
\newcommand{\Unode}{\mbox{``\textsf{U}''}}

\newcommand{\A}{\mathcal{A}}
\newcommand{\B}{\mathcal{B}}

\newcommand{\E}{\mathcal{E}}
\newcommand{\F}{\mathcal{F}}

\renewcommand{\O}{\mathcal{O}}

\renewcommand{\S}{\mathcal{S}}

\def\StrPt{\SRules\cup\Facts}
\def\kitSink{\StrPt\cup\DD}

\begin{document}

\title{An Argumentation-Based Framework to Address the \\ Attribution Problem in Cyber-Warfare}

\author{Paulo Shakarian$^{1}$ \ \ Gerardo I. Simari$^{2}$ \ \ Geoffrey Moores$^{1}$ \ \
Simon Parsons$^{3}$ \ \ Marcelo A. Falappa$^{4}$ \\[1ex]
$^{1}$Dept.\ of Electrical Engineering and Computer Science, U.S.\ Military Academy, West Point, NY\\
$^{2}$Dept.\ of Computer Science, University of Oxford, Oxford, UK\\
$^{3}$Dept.\ of Computer Science, University of Liverpool, Liverpool, UK\\
$^{4}$Dep.\ de Cs.\ e Ing.\ de la Computaci\'on, Univ.\ Nac.\ del Sur, Bah\'{\i}a Blanca, Argentina and CONICET \\
paulo@shakarian.net, gerardo.simari@cs.ox.ac.uk, geoffrey.moores@usma.edu\\ s.d.parsons@liverpool.ac.uk, mfalappa@cs.uns.edu.ar
}

\maketitle

\begin{abstract}
Attributing a cyber-operation through the use of multiple pieces of technical evidence (i.e., malware reverse-engineering
and source tracking) and conventional intelligence sources (i.e., human or signals intelligence) is a difficult problem
not only due to the effort required to obtain evidence, but the ease with which an adversary can plant false evidence.
In this paper, we introduce a formal reasoning system called the InCA (Intelligent Cyber Attribution) framework that
is designed to aid an analyst in the attribution of a cyber-operation even when the available information is conflicting
and/or uncertain.  Our approach combines argumentation-based reasoning, logic programming, and probabilistic models to not only attribute an operation but also explain to the analyst why the system reaches its conclusions.
\end{abstract}

\section{Introduction}
\label{sec:intro}

An important issue in cyber-warfare is the puzzle of determining who was responsible for a given cyber-operation --
be it an incident of attack, reconnaissance, or information theft.  This is known as the ``attribution
problem''~\cite{shakBook13}.  The difficulty of this problem stems not only from the amount of effort required to find
forensic clues but also the ease with which an attacker can plant false clues to mislead security personnel.
Further, while techniques such as forensics and reverse-engineering~\cite{alth13}, source
tracking~\cite{thonnard10}, honeypots~\cite{spi03}, and sinkholing~\cite{shadow10} are commonly employed to find evidence
that can lead to attribution, it is unclear how this evidence is to be combined and reasoned about.
In a military setting, such evidence is augmented with normal intelligence collection, such as human intelligence (HUMINT),
signals intelligence (SIGINT) and other means -- this adds additional complications to the task of attributing a given
operation.
Essentially, cyber-attribution is a highly-technical intelligence analysis problem where an analyst must consider a variety
of sources, each with its associated level of confidence, to provide a decision maker (e.g., a military commander) insight
into who conducted a given operation.

As it is well known that people's ability to conduct intelligence analysis is limited~\cite{Heuer07}, and due to the highly
technical nature of many cyber evidence-gathering techniques, an automated reasoning system would be best suited for
the task.
Such a system must be able to accomplish several goals, among which we distinguish the following main capabilities:

\begin{enumerate}
\item Reason about evidence in a formal, principled manner, i.e., relying on strong mathematical foundations.

\item Consider evidence for cyber attribution associated with some level of probabilistic uncertainty.

\item 
Consider logical rules that allow for the system to draw conclusions based on certain pieces of evidence and iteratively apply such rules.

\item 
Consider pieces of information that may not be compatible with each other, decide which information is most relevant, and express why.

\item  
Attribute a given cyber-operation based on the above-described features and provide the analyst with the ability to understand how the system arrived at that conclusion.
\end{enumerate}

In this paper we present the InCA (Intelligent Cyber Attribution) framework, which meets all of the above qualities.
Our approach relies on several techniques from the artificial intelligence community, including argumentation, logic
programming, and probabilistic reasoning.  
We first outline the underlying mathematical framework and provide examples
based on real-world cases of cyber-attribution (cf.\ Section~\ref{prelims}); then, 
in Sections~\ref{sec:probarg} and~\ref{sec:attrib-queries}, we formally present InCA and attribution queries, respectively.
Finally, we discuss conclusions and future work in Section~\ref{sec:conc}.

\section{Two Kinds of Models}
\label{prelims}

Our approach relies on \textit{two separate models of the world}.  The first, called the
\textit{\textbf{environmental model}} (EM) is used to describe the background knowledge and is probabilistic in nature.
The second one, called the \textit{\textbf{analytical model}} (AM) is used to analyze competing hypotheses that can account
for a given phenomenon (in this case, a cyber-operation).  The EM \textit{must be consistent} -- this simply means that
there must
exist a probability distribution over the possible states of the world that satisfies all of the constraints in the model,
as well as the axioms of probability theory.
On the contrary, the AM will allow for contradictory information as the system must have the capability to reason about competing explanations for a given cyber-operation.  In general, the EM contains knowledge such as evidence, intelligence reporting, or knowledge about actors, software, and systems.  The AM, on the other hand, contains ideas the analyst concludes based on the information in the EM.
Figure~\ref{emAmTable} gives some examples of the types of information in the two models.
Note that an analyst (or automated system) could assign a probability to statements in the EM column whereas
statements in the AM column can be true or false depending on a certain combination (or several possible combinations) of statements from the EM.
We now formally describe these two models as
well as a technique for \textit{annotating} knowledge in the AM with
information from the EM -- these annotations specify the conditions
under which the various statements in the AM can potentially be true.

\begin{figure}
\centering
\footnotesize
\begin{tabular}{ll}
\hline
\textbf{EM} & \textbf{AM} \\\hline
``Malware X was compiled & ``Malware X was compiled \\
 on a system using the   & on a system in English- \\
English language.''      & speaking country Y.''\\
\hline
``Malware W and malware X & ``Malware W and \\
were created in a similar & malware X are \\
coding style.''           & related.''\\
\hline
``Country Y and country Z & ``Country Y has a motive to \\
are currently at war.''  & launch a cyber-attack against \\
                         & country Z.'' \\
\hline
``Country Y has a significant  & ``Country Y has the capability \\
investment in math-science-    & to conduct a cyber-attack.'' \\
engineering (MSE) education.'' & \\
\hline
\end{tabular}
\caption{Example observations -- EM vs.\ AM.}
\label{emAmTable}
\end{figure}

Before describing the two models in detail, we first introduce the language used to describe them.
Variable and constant symbols represent items such as computer systems, types of cyber operations, actors (e.g., nation states,
hacking groups), and other technical and/or intelligence information.  The set of all variable symbols is denoted with
$\Var$, and the set of all constants is denoted with $\Cons$.  For our framework, we shall require two subsets of $\Cons$,
$\actors$ and $\ops$, that specify the actors that could conduct cyber-operations and the operations themselves, respectively.
In the examples in this paper, we will use capital letters to represent variables (e.g., $X,Y,Z$). The constants in $\actors$ and $\ops$ that we use in the running example are specified in the following example.

\begin{example}
The following (fictitious) actors and cyber-operations will be used in our examples:
\begin{eqnarray}
\actors &=& \{\baja,\kras,\mojave\} \\
\ops    &=& \{\worm \}
\end{eqnarray}
\smallbsq
\end{example}

The next component in the model is a set of predicate symbols.  These constructs can accept zero or more variables or constants as
arguments, and map to either {\em true} or {\em false}.  \textit{Note that the EM and AM use separate sets of predicate
symbols} -- however, they can share variables and constants.  The sets of predicates for the EM and AM are denoted with
$\PredE,\PredA$, respectively.  In InCA, we require $\PredA$ to include the binary predicate $\condOp(X,Y)$, where $X$ is an
actor and $Y$ is a cyber-operation.  Intuitively, this means that actor $X$ conducted operation $Y$.  For instance,
$\condOp(\baja,\worm)$ is true if $\baja$ was responsible for cyber-operation $\worm$.
A sample set of predicate symbols for the analysis of a cyber attack between two states over contention of a particular
industry is shown in Figure~\ref{fig:runEx}; these will be used in examples throughout the paper.

\begin{figure*}
\centering
\fbox{
\parbox{\textwidth}{
\begin{tabular}{lll}
$\PredE$: & $\origIP(M,X)$ & Malware $M$ originated from an IP address belonging to actor $X$.\\
& $\mwInOp(M,O)$ & Malware $M$ was used in cyber-operation $O$.\\
& $\mwHint(M,X)$ & Malware $M$ contained a hint that it was created by actor $X$.\\
& $\compiledLan(M,C)$ & Malware $M$ was compiled in a system that used language $C$.\\
& $\natLang(X,C)$ & Language $C$ is the native language of actor $X$.\\
& $\inLgConf(X,X')$ & Actors $X$ and $X'$ are in a larger conflict with each other.\\
& $\numTtMse(X,N)$ & There are at least $N$ number of top-tier math-science-engineering universities in country $X$.\\
& $\infSys(X,M)$ & Systems belonging to actor $X$ were infected with malware $M$.\\
& $\cybCapAge(X,N)$ & Actor $X$ has had a cyber-warfare capability for $N$ years or less.\\
& $\govCybCap(X)$ & Actor $X$ has a government cyber-security lab.\\ [6pt]
$\PredA$: & $\condOp(X,O)$ & Actor $X$ conducted cyber-operation $O$.\\
 & $\evidOf(X,O)$ & There is evidence that actor $X$ conducted cyber-operation $O$.\\
& $\motiv(X,X')$ & Actor $X$ had a motive to launch a cyber-attack against actor $X'$.\\
& $\isCap(X,O)$ & Actor $X$ is capable of conducting cyber-operation $O$.\\
& $\tgt(X,O)$ & Actor $X$ was the target of cyber-operation $O$.\\
& $\hasMse(X)$ & Actor $X$ has a significant investment in math-science-engineering education.\\
& $\expCw(X)$ & Actor $X$ has experience in conducting cyber-operations.
\end{tabular}
}}
\caption{Predicate definitions for the environment and analytical models in the running example.}
\label{fig:runEx}
\end{figure*}

A construct formed with a predicate and constants as arguments is known as a \textit{ground atom}
(we shall often deal with ground atoms).  The sets of all ground atoms for EM and AM
are denoted with $\GndE$ and $\GndA$, respectively.

\begin{example}
The following are examples of ground atoms over the predicates given in Figure~\ref{fig:runEx}.
\begin{eqnarray*}
\GndE: && \origIP(\mwsam,\kras),\\
       && \mwHint(\mwsam,\kras),\\
       && \inLgConf(\kras,\baja), \\
       && \numTtMse(\kras,2)\\[2ex]
\GndA: && \evidOf(\mojave,\worm), \\
       && \motiv(\baja,\kras),\\
       && \expCw(\baja),\\
       && \tgt(\kras,\worm)
\end{eqnarray*}
\smallbsq
\end{example}

\noindent
For a given set of ground atoms, a \textit{world} is a subset of the atoms that are considered to be true
(ground atoms not in the world are false).  Hence, there are $2^{|\GndE|}$ possible worlds in the
EM and $2^{|\GndA|}$ worlds in the AM, denoted with $\wldE$ and $\wldA$, respectively.

Clearly, even a moderate number of ground atoms can yield an enormous number of worlds to explore.  One way to reduce
the number of worlds is to include \textit{integrity constraints}, which allow us to eliminate certain worlds from
consideration -- they simply are not possible in the setting being modeled.  Our principle integrity constraint will be
of the form:
\[
\textsf{oneOf}(\cala')
\]
where $\cala'$ is a subset of ground atoms.
Intuitively, this says that any world where more than one of the atoms from set $\cala'$ appear is invalid.
Let $\ICE$ and $\ICA$ be the sets of integrity constraints for the EM and AM,
respectively, and the sets of worlds that conform to these constraints be $\wldE(\ICE),\wldA(\ICA)$, respectively.

Atoms can also be combined into formulas using standard logical connectives: conjunction ({\em and}),
disjunction ({\em or}), and negation ({\em not}).  These are written using the symbols $\wedge,\vee,\neg$, respectively.
We say a world ($w$) \textit{satisfies} a formula ($f$), written $w\models f$, based on the following inductive definition:
\begin{itemize}
\item if $f$ is a single atom, then $w \models f$ iff $f \in w$;

\item if $f = \neg f'$ then $w \models f$ iff $w \not\models f'$;

\item if $f=f' \wedge f''$ then $w \models f$ iff $w\models f'$ and $w\models f''$; and

\item if $f=f' \vee f''$ then $w \models f$ iff $w \models f'$ or $w \models f''$.
\end{itemize}
We use the notation $\fEM, \fAM$ to denote the set of all possible (ground) formulas in the EM and AM, respectively.
Also, note that we use the notation $\top, \bot$ to represent tautologies (formulas that are true in all worlds) and  contradictions (formulas that are false in all worlds), respectively.

\subsection{Environmental Model}
\label{sec:em}

In this section we describe the first of the two models, namely the EM or environmental model.
This model is largely based on the probabilistic logic of~\cite{nil86}, which we now briefly review.

First, we define a \textit{probabilistic formula} that consists of a formula $f$ over atoms from $\GndE$, a real number
$p$ in the interval $[0,1]$, and an error tolerance $\epsilon \in [0,\min(p,1-p)]$.  A probabilistic formula is written as:
$f : p \pm \epsilon$.
Intuitively, this statement is interpreted as ``formula $f$ is true with probability between $p - \epsilon$
and $p + \epsilon$'' -- note that we make no statement about the probability distribution over this interval.
The uncertainty regarding the probability values stems from the fact
that certain assumptions (such as probabilistic independence) may not be suitable in the environment being modeled.

\begin{example}
\label{logPgmEx}
To continue our running example, consider the following set $\kE$:
\begin{eqnarray*}
f_1 &=& \govCybCap(\baja): 0.8 \pm 0.1\\
f_2 &=& \cybCapAge(\baja,5) : 0.2 \pm 0.1\\
f_3 &=& \numTtMse(\baja, 2) : 0.8 \pm 0.1 \\
f_4 &=& \mwHint(\mwsam,\mojave)\\
&& \mbox{}\wedge \compiledLan(\worm,\eng) : 0.7 \pm 0.2\\
f_5 &=& \mwInOp(\mwsam,\worm) \\
&&\mbox{}\wedge \mwRel(\mwsam,\mwsamT) \\
&&\mbox{}\wedge \mwHint(\mwsamT,\mojave) : 0.6 \pm 0.1\\
f_6 &=& \inLgConf(\baja,\kras)\\
&& \mbox{}\vee \neg\cooper(\baja,\kras) : 0.9 \pm 0.1\\
f_7 &=& \origIP(\mwsam,\baja) : 1 \pm 0
\end{eqnarray*}
Throughout the paper, let $\kE' = \{f_1,f_2,f_3\}$.
\smallbsq
\end{example}

We now consider a probability distribution $\Pr$ over the set $\wldE(\ICE)$.  We say that $\Pr$ \textit{satisfies} probabilistic formula $f:p \pm \epsilon$ iff the following holds:
$
p - \epsilon \leq \sum_{w \in \wldE(\ICE)}\Pr(w) \leq p+\epsilon.
$
A set $\kE$ of probabilistic formulas is called a \textit{knowledge base}.  We say that a probability
distribution over $\wldE(\ICE)$ \emph{satisfies} $\kE$ if and only if it satisfies all probabilistic formulas in $\kE$.

It is possible to create probabilistic knowledge bases for which there is no satisfying probability distribution.
The following is a simple example of this:
\begin{eqnarray*}
\lefteqn{\condOp(\kras,\worm)}&& \\
&& \mbox{} \vee \condOp(\baja,\worm) : 0.4 \pm 0;
\\
\lefteqn{\condOp(\kras,\worm)}&& \\
&& \mbox{} \wedge \condOp(\baja,\worm) : 0.6 \pm 0.1.
\end{eqnarray*}

Formulas and knowledge bases of this sort are \textit{inconsistent}.  In this paper, we assume that information
is properly extracted from a set of historic data and hence consistent; (recall that inconsistent information can only be handled in the AM, not the EM). A consistent
knowledge base could also be obtained as a result of curation by experts, such that all inconsistencies were removed --
see \cite{amai07,apt1} for algorithms for learning rules of this type.

The main kind of query that we require for the probabilistic model is the \textit{maximum entailment} problem:
given a knowledge base $\kE$ and a (non-probabilistic) formula $q$, identify $p,\epsilon$ such that all valid probability
distributions $\Pr$ that satisfy $\kE$ also satisfy $q: p \pm \epsilon$, and there does not exist $p', \epsilon'$ s.t.\
$[p-\epsilon,p+\epsilon] \supset [p'-\epsilon',p'+\epsilon']$, where all probability distributions $\Pr$ that satisfy
$\kE$ also satisfy $q : p' \pm \epsilon'$.
That is, given $q$, can we determine the probability (with maximum tolerance) of statement $q$ given the
information in $\kE$?
The approach adopted in~\cite{nil86} to solve this problem works as follows. First, we must solve the linear program defined next.

\begin{definition}[\textsf{EM-LP-MIN}]
\label{def:em-lp-min}
Given a knowledge base $\kE$ and a formula $q$:
\begin{itemize}
\item create a variable $x_i$ for each $w_i \in \wldE(\ICE)$;

\item for each $f_j : p_j \pm \epsilon_j \in \kE$, create constraint:
\[
p_j - \epsilon_j \leq \sum_{w_i \in \wldE(\ICE) \textit{ s.t.\ } w_i \models f_j}x_i \leq p_j + \epsilon_j;
\]

\item
finally, we also have a constraint:
\[
\sum_{w_i \in \wldE(\ICE)} x_i = 1.
\]
\end{itemize}

\noindent
The objective is to minimize the function:
\[
\sum_{w_i \in \wldE(\ICE) \textit{ s.t.\ } w_i \models q} x_i.
\]
We use the notation $\textsf{EP-LP-MIN}(\kE,q)$ to refer to the value of the objective function in the
solution to the \textsf{EM-LP-MIN} constraints.
\end{definition}

Let $\ell$ be the result of the process described in Definition~\ref{def:em-lp-min}.  
The next step is to solve the linear program a second time, but
instead maximizing the objective function (we shall refer to this as \textsf{EM-LP-MAX}) -- let $u$ be the result of this
operation.
In \cite{nil86}, it is shown that $\epsilon = \frac{u-\ell}{2}$ and $p = \ell+\epsilon$ is the solution to the maximum entailment problem.  We note that although the above linear program has an exponential number of variables in the worst
case (i.e., no integrity constraints), the presence of constraints has the potential to greatly reduce this space.
Further, there are also good heuristics (cf.\ \cite{amai07,amai12}) that have been shown to provide highly accurate
approximations with a reduced-size linear program.

\begin{example}
\label{linPrgEx}
Consider KB $\kE'$ from Example~\ref{logPgmEx} and a set of ground atoms restricted to those that appear in that program.  
Hence, we have:
\begin{eqnarray*}
w_1 &=& \{ \govCybCap(\baja), \cybCapAge(\baja, 5),\\
&& \numTtMse(\baja, 2) \}\\
w_2 &=& \{ \govCybCap(\baja), \cybCapAge(\baja, 5) \}\\
w_3 &=& \{ \govCybCap(\baja), \numTtMse(\baja, 2) \}\\
w_4 &=& \{ \cybCapAge(\baja, 5), \numTtMse(\baja, 2) \}\\
w_5 &=& \{ \cybCapAge(\baja, 5) \}\\
w_6 &=& \{ \govCybCap(\baja) \}\\
w_7 &=& \{ \numTtMse(\baja, 2) \}\\
w_8 &=& \emptyset
\end{eqnarray*}
and suppose we wish to compute the probability for formula:
\[
q = \govCybCap(\baja) \vee \numTtMse(\baja, 2).
\]
For each formula in $\kE$ we have a constraint, and for each world above
we have a variable.  An objective function is created based on the worlds that satisfy the query formula
(here, worlds $w_1$--$w_4$, $w_6$, $w_7$).  Hence, $\textsf{EP-LP-MIN}(\kE',q)$ can be written as:
\begin{eqnarray*}
\max & x_1+x_2+x_3+x_4+x_6+x_7 & \mathit{w.r.t.:}\\
0.7\leq & x_1+x_2+x_3+x_6 & \leq 0.9\\
0.1 \leq & x_1+x_2+x_4+x_5 & \leq 0.3\\
0.8 \leq & x_1+x_3+x_4+x_7 & \leq 1\\
& x_1+x_2+x_3+x_4+x_5+x_6+x_7+x_8 & = 1
\end{eqnarray*}
We can now solve $\textsf{EP-LP-MAX}(\kE',q)$ and
$\textsf{EP-LP-MIN}(\kE',q)$ to get solution $0.9 \pm 0.1$.
\smallbsq
\end{example}

\subsection{Analytical Model}
\label{sec:am}

For the analytical model (AM), we choose a structured argumentation framework~\cite{RahwanSimari2009} due to
several characteristics that make such frameworks highly applicable to cyber-warfare domains.  Unlike the EM, which describes
probabilistic information about the state of the real world, the AM must allow for competing ideas --
it \textit{must be able to represent contradictory information}. The algorithmic approach
allows for the creation of \textit{arguments} based on the AM that may ``compete'' with each other to describe who
conducted a given cyber-operation.  In this competition -- known as a \textit{dialectical process} -- one argument may
defeat another based on a \textit{comparison criterion} that determines the prevailing argument.  Resulting from this
process, the InCA framework will determine arguments that are \textit{warranted} (those that are not \emph{defeated} by other
arguments) thereby providing a suitable explanation for a given cyber-operation.

The transparency provided by the system can allow analysts to identify potentially incorrect input information and fine-tune
the models or, alternatively, collect more information.  In short, argumentation-based reasoning has been studied as a natural way
to manage a set of inconsistent information -- it is the way humans settle disputes. As we will see, another desirable
characteristic of (structured) argumentation frameworks is that, once a conclusion is reached, we are left with an explanation
of how we arrived at it and information about why a given argument is warranted; this is very important information for analysts to have.  In this section, we recall some preliminaries of the underlying argumentation framework used, and then
introduce the analytical model (AM).

\subsubsection*{Defeasible Logic Programming with Presumptions}
\label{sec:delp}

\DLP\ with Presumptions (\PDeLP)~\cite{MartinezGS12} is a formalism combining Logic Programming
with Defeasible Argumentation. We now briefly recall the basics of \PDeLP;
we refer the reader to~\cite{Garcia-Simari04-TPLP,MartinezGS12} for the complete presentation.
The formalism contains several different constructs: facts, presumptions, strict rules, and defeasible rules.
Facts are statements about the analysis that can always be considered to be true, while presumptions are statements
that may or may not be true.  Strict rules specify logical consequences of a set of facts or presumptions (similar to an implication, though not the same) that must always occur, while defeasible rules specify logical consequences that may be assumed to be true when
no contradicting information is present.  These constructs are used in the construction of \emph{arguments},
and are part of a \PDeLP\ program, which is a set of facts, strict rules, presumptions, and defeasible rules.
Formally, we use the notation $\PP = \pdelpprog$ to denote a \PDeLP\ program, where $\SRules$ is the
set of strict rules, $\Facts$ is the set of facts, $\DRules$ is the set of defeasible rules, and $\Presumps$ is the set
of presumptions.  In Figure~\ref{fig:gndArgEx}, we provide an example $\PP$.
We now describe each of these constructs in detail.

\medskip
\noindent
\textbf{Facts} ($\Facts$) are ground literals representing atomic information or its negation, using
strong negation ``\no''.  Note that all of the literals in our framework must be formed with a predicate from the
set $\PredA$.  Note that information in this form cannot be contradicted.

\medskip
\noindent
\textbf{Strict Rules} ($\SRules$) represent non-defeasible cause-and-effect information that resembles an implication (though the
semantics is different since the contrapositive does not hold) and are of the form
$
\srule{L_0}{L_1, \ldots, L_n}
$,
where  $L_0$ is a ground literal and $\{L_i\}_{i>0}$ is a set of ground literals.

\medskip
\noindent
\textbf{Presumptions} ($\Presumps$) are ground literals of the same form as facts, except that they are not taken as being true
but rather defeasible, which means that they can be contradicted. Presumptions are denoted in the same manner as facts,
except that the symbol $\defleftarrow$ is added.  While any literal can be used as a presumption in InCA, we specifically
require all literals created with the predicate $\condOp$ to be defeasible.

\medskip
\noindent
\textbf{Defeasible Rules} ($\DRules$) represent tentative knowledge that can be used if nothing can be posed against it.
Just as presumptions are the defeasible counterpart of facts, defeasible rules are the defeasible counterpart of
strict rules. They are of the form
$
\drule{L_0}{L_1, \ldots, L_n}
$,
where  $L_0$ is a ground literal and $\{L_i\}_{i>0}$ is a set of ground literals.  Note that with both strict and
defeasible rules, {\it strong negation} is allowed in the head of rules, and hence may be used to represent contradictory
knowledge.

Even though the above constructs are ground, we allow for schematic versions with variables that are used to represent
sets of ground rules.
We denote variables with strings starting with an uppercase letter;
Figure~\ref{fig:ngArgEx} shows a non-ground example.

When a cyber-operation occurs, InCA must derive arguments as to who could have potentially conducted the action.
Derivation follows the same mechanism of Logic Programming~\cite{ll87}. Since rule heads can contain strong
negation, it is possible to defeasibly derive contradictory literals from a program.
For the treatment of contradictory knowledge, \PDeLP\ incorporates a defeasible argumentation formalism
that allows the identification of the pieces of knowledge that are in conflict,
and through the previously mentioned \emph{dialectical process} decides which information prevails as warranted.

\begin{figure}
\fbox{
\parbox{0.96\columnwidth}{
\begin{tabular}{lll}
$\Facts:$& $\f_{1a}=$ & $\evidOf(\baja,\worm)$\\
& $\f_{1b}=$ & $\evidOf(\mojave,\worm)$\\
& $\f_2=$ & $\motiv(\baja,\kras)$\\
\multicolumn{3}{c}{\rule{0.9\columnwidth}{0.4pt}} \\
$\SRules:$& $\r_{1a}=$ & $\neg\condOp(\baja,\worm) \leftarrow$ \\
          &            &  \hskip6mm $\condOp(\mojave,\worm)$\\
& $\r_{1b}=$ & $\neg\condOp(\mojave,\worm)\leftarrow$ \\
&            & \hskip6mm $\condOp(\baja,\worm)$\\
&$\r_{2a}=$ & $\condOp(\baja,\worm) \leftarrow$ \\
&           & \hskip6mm $\evidOf(\baja,\worm),$ \\
&           & \hskip6mm $\isCap(\baja,\worm),$ \\
&           & \hskip6mm $\motiv(\baja,\kras),$\\
&           & \hskip6mm $\tgt(\kras,\worm)$\\
&$\r_{2b}=$ & $\condOp(\mojave,\worm) \leftarrow$ \\
&           & \hskip6mm $\evidOf(\mojave,\worm),$ \\
&           & \hskip6mm $\isCap(\mojave,\worm),$\\
&           & \hskip6mm $\motiv(\mojave,\kras),$ \\
&           & \hskip6mm  $\tgt(\kras,\worm)$\\
\multicolumn{3}{c}{\rule{0.9\columnwidth}{0.4pt}} \\
$\Presumps:$& $\p_1=$&  $\hasMse(\baja) \; \defleftarrow$\\
&$\p_2=$&$\tgt(\kras,\worm) \; \defleftarrow$\\
&$\p_3=$&$\neg\expCw(\baja) \; \defleftarrow$\\
\multicolumn{3}{c}{\rule{0.9\columnwidth}{0.4pt}} \\
$\DRules:$& $\d_{1a}=$ & $\condOp(\baja,\worm) \; \defleftarrow$ \\
          &            & \hskip6mm  $\evidOf(\baja,\worm)$\\
          & $\d_{1b}=$ & $\condOp(\mojave,\worm) \; \defleftarrow$ \\
          &            & \hskip6mm  $\evidOf(\mojave,\worm)$\\
          & $\d_2=$    & $\condOp(\baja,\worm) \; \defleftarrow$ \\
          &            & \hskip6mm  $\isCap(\baja,\worm)$\\
          & $\d_3=$    & $\condOp(\baja,\worm) \; \defleftarrow$ \\
          &            & \hskip6mm  $\motiv(\baja,\kras),$ \\
          &            & \hskip6mm  $\tgt(\kras,\worm)$ \\
          & $\d_4=$    & $\isCap(\baja,\worm) \; \defleftarrow$\\
					&						 & \hskip6mm  $\hasMse(\baja)$\\
          & $\d_{5a}=$ & $\neg\isCap(\baja,\worm) \; \defleftarrow \neg\expCw(\baja)$\\
          & $\d_{5b}=$ & $\neg\isCap(\mojave,\worm) \; \defleftarrow $\\
					&						 & \hskip6mm  $\neg\expCw(\mojave)$
\end{tabular}
}}
\caption{A ground argumentation framework.}
\label{fig:gndArgEx}
\end{figure}

\begin{figure}
\fbox{
\parbox{0.96\columnwidth}{
\begin{tabular}{lll}
$\Facts:$& $\f_1=$ & $\evidOf(\baja,\worm)$\\
& $\f_2=$ & $\motiv(\baja,\kras)$\\
\multicolumn{3}{c}{\rule{0.9\columnwidth}{0.4pt}} \\
$\SRules:$& $\r_1=$ & $\neg\condOp(X,O)\leftarrow \condOp(X',O),$\\
&				 & \hskip6mm $X \neq X'$\\
&$\r_2=$ & $\condOp(X,O) \leftarrow \evidOf(X,O),$ \\
&        & \hskip6mm $\isCap(X,O),\motiv(X,X'), $\\
&&\hskip6mm$\tgt(X',O),X \neq X'$\\
\multicolumn{3}{c}{\rule{0.9\columnwidth}{0.4pt}} \\
$\Presumps:$& $\p_1=$&  $\hasMse(\baja) \; \defleftarrow$\\
&$\p_2=$&$\tgt(\kras,\worm) \; \defleftarrow$\\
&$\p_3=$&$\neg\expCw(\baja) \; \defleftarrow$\\
\multicolumn{3}{c}{\rule{0.9\columnwidth}{0.4pt}} \\
$\DRules:$& $\d_1=$ & $\condOp(X,O) \;  \defleftarrow \evidOf(X,O)$\\
& $\d_2=$ & $\condOp(X,O) \;  \defleftarrow \isCap(X,O)$\\
& $\d_3=$ & $\condOp(X,O)  \; \defleftarrow \motiv(X,X'), \tgt(X',O)$\\
& $\d_4=$ & $\isCap(X,O) \;  \defleftarrow \hasMse(X)$\\
& $\d_5=$ & $\neg\isCap(X,O) \;  \defleftarrow \neg\expCw(X)$
\end{tabular}
}}
\caption{A non-ground argumentation framework.}
\label{fig:ngArgEx}
\end{figure}

This dialectical process involves the construction and evaluation
of arguments that either support or interfere with a given query, building a \emph{dialectical tree}
in the process.  Formally, we have:

\begin{definition}[Argument]
An \emph{argument} $\langle\Arg, L\rangle$ for a literal $L$ is a pair of the literal and a (possibly empty) set of the EM ($\Arg \subseteq \PP$) that provides a minimal proof for $L$ meeting the requirements: (1.) $L$ is defeasibly derived from $\Arg$, (2.) $\StrPt\cup\Arg$ is not contradictory, and (3.) $\Arg$ is a minimal subset of $\DD$ satisfying 1 and 2, denoted \AL.

Literal $L$ is called the {\em conclusion} supported by the argument, and $\Arg$ is the \emph{support} of the argument. An argument $\langle \B, L\rangle$
is a {\em subargument} of $\langle \A, L'\rangle$ iff $\B \subseteq \A$.  An argument $\langle\A, L\rangle$ is {\em presumptive} iff $\A \cap \Presumps$ is not empty.  We will also use $\SRules(\A) = \A \cap \SRules$, $\Facts(\A) = \A \cap \Facts$, $\DRules(\A) = \A \cap \DRules$, and $\Presumps(\A) = \A \cap \Presumps$.
\end{definition}

Note that our definition differs slightly from that of \cite{Simari-Loui92} where DeLP is introduced, as we include strict rules and facts as part of the argument.  The reason for this will become clear in Section~\ref{sec:probarg}.
Arguments for our scenario are shown in the following example.

\begin{example}
\label{ex:args}
Figure~\ref{fig:gndArgsEx} shows example arguments based on the knowledge base from Figure~\ref{fig:gndArgEx}. 
Note that the following relationship exists:
\begin{quote}
$\<\cala_5,\isCap(\baja,\worm)\>$ is a sub-argument of\\ 
$\<\cala_2,\condOp(\baja,\worm)\>$ and \\ 
$\<\cala_3,\condOp(\baja,\worm)\>$.
\smallbsq
\end{quote}
\end{example}

\begin{figure}
\fbox{
\parbox{0.96\columnwidth}{
\begin{tabular}{ll}
$\<\cala_1,\condOp(\baja,\worm)\>$& $\cala_1 = \{\f_{1a},\d_{1a}\}$\\
$\<\cala_2,\condOp(\baja,\worm)\>$& $\cala_2 = \{\p_1,\p_2,\d_4,\r_{2a},$\\
&\hskip8mm $\f_{1a},\f_2\}$\\
$\<\cala_3,\condOp(\baja,\worm)\>$& $\cala_3 = \{\p_1,\d_2,\d_4\}$\\
$\<\cala_4,\condOp(\baja,\worm)\>$& $\cala_4 = \{\p_2,\d_3,\f_2\}$\\
$\<\cala_5,\isCap(\baja,\worm)\>$& $\cala_5 = \{\p_1,\d_4\}$\\
$\<\cala_6,\neg\condOp(\baja,\worm)\>$& $\cala_6 = \{\d_{1b},\f_{1b},\r_{1a}\}$\\
$\<\cala_7,\neg\isCap(\baja,\worm)\>$& $\cala_7 = \{\p_{3},\d_{5a}\}$
\end{tabular}
}}
\caption{Example ground arguments from Figure~\ref{fig:gndArgEx}.}
\label{fig:gndArgsEx}
\end{figure}

Given argument $\langle \A_1, L_1\rangle$, counter-arguments are arguments that contradict it. Argument
$\langle \A_2, L_2\rangle$ {\em counterargues} or {\em attacks} $\langle \A_1, L_1\rangle$ literal $L'$ iff there
exists a subargument $\langle \A, L''\rangle$ of $\langle \A_1, L_1\rangle$ s.t.\ set
$\SRules(\A_1) \cup \SRules(\A_2) \cup \Facts(\A_1) \cup \Facts(\A_2) \cup \{L_2,L''\}$ is contradictory.

\begin{example}
Consider the arguments from Example~\ref{ex:args}. The following are some of the
attack relationships between them:
$\cala_1$, $\cala_2$, $\cala_3$, and $\cala_4$ all attack $\cala_6$;
$\cala_5$ attacks $\cala_7$; and
$\cala_7$ attacks $\cala_2$.
\smallbsq
\end{example}

A {\em proper defeater} of an argument $\langle A, L\rangle$ is a counter-argument
that -- by some criterion -- is considered to be better than $\langle A, L\rangle$; if the two are incomparable according
to this criterion, the counterargument is said to be a {\em blocking} defeater.
An important characteristic of \PDeLP\ is that the argument comparison criterion is modular, and thus the most appropriate criterion for the domain that is being represented can be selected; the default criterion used in classical defeasible logic programming (from which \PDeLP\ is derived) is \emph{generalized specificity}~\cite{jncl2003}, though an extension of this criterion is required for arguments using presumptions~\cite{MartinezGS12}. We briefly recall this criterion next -- the first
definition is for generalized specificity, which is subsequently used in the definition of presumption-enabled
specificity.

\begin{definition}
\label{def:pspecificity}
Let $\Prog = \pdelpprog$ be a \PDeLP\ program and let $\F$ be the set of all literals that
have a defeasible derivation from $\Prog$.
An argument
$\langle \A_1, L_1\rangle$ is \emph{preferred to} $\langle \A_2, L_2\rangle$, denoted with
$\A_1 \succ_{PS} \A_2$ iff the two following conditions hold:

\begin{enumerate}
\item For all $H \subseteq \F$, $\SRules(\cala_1)\cup\SRules(\A_2) \cup H$ is non-contradictory: if there is a derivation for $L_1$
from $\SRules(\A_2)\cup \SRules(\A_1)\cup\DRules(\A_1) \cup H$,
and there is no derivation for $L_1$ from $\SRules(\A_1) \cup \SRules(\A_2) \cup H$, then there is a derivation for $L_2$ from
$\SRules(\A_1) \cup \SRules(\A_2) \cup \DRules(\A_2) \cup H$.

\item There is at least one set $H' \subseteq \F$, $\SRules(\cala_1)\cup\SRules(\A_2) \cup H'$ is non-contradictory, such 
that there is a derivation for $L_2$ from $\SRules(\cala_1)\cup\SRules(\A_2) \cup H' \cup \DRules(\A_2)$, there is no 
derivation for $L_2$ from $\SRules(\cala_1)\cup\SRules(\A_2)  \cup H'$, and
there is no derivation for $L_1$ from $\SRules(\cala_1)\cup\SRules(\A_2)  \cup H' \cup \DRules(\A_1)$.
\end{enumerate}
\end{definition}

Intuitively, the principle of specificity says that, in the presence of two conflicting lines of argument about a proposition, the one that uses more of the available information is more convincing. A classic example involves a bird, Tweety, and
arguments stating that it both flies (because it is a bird) and doesn't fly (because it is a penguin).
The latter argument uses more information about Tweety -- it is more specific -- and is thus the stronger of the two.

\begin{definition}[\cite{MartinezGS12}]
\label{def:MGEPref}
Let $\Prog = \pdelpprog$ be a \PDeLP\ program. An argument
$\langle \A_1, L_1\rangle$ is \emph{preferred to} $\langle \A_2, L_2\rangle$, denoted with
$\A_1 \succ \A_2$ iff any of the following conditions hold:

\begin{enumerate}
\item $\langle \A_1, L_1\rangle$ and $\langle \A_2, L_2\rangle$ are both factual arguments and
$\langle \A_1, L_1\rangle \succ_{PS} \langle \A_2, L_2\rangle$.

\item $\langle \A_1, L_1\rangle$ is a factual argument and $\langle \A_2, L_2\rangle$ is a presumptive argument.

\item $\langle \A_1, L_1\rangle$ and $\langle \A_2, L_2\rangle$ are presumptive arguments, and

\begin{enumerate}
\item $\neg(\Presumps(\A_1) \subseteq \Presumps(\A_2))$, or \label{itema}

\item $\Presumps(\A_1) = \Presumps(\A_2)$ and $\langle \A_1, L_1\rangle \succ_{PS} \langle \A_2, L_2\rangle$.\label{itemb}
\end{enumerate}
\end{enumerate}
\end{definition}
Generally, if $\A, \B$ are arguments with rules $X$ and $Y$, resp., and $X \subset Y$, then $\A$
is stronger than $\B$. This also holds when $\A$ and $\B$ use presumptions $P_1$ and $P_2$, resp.,
and $P_1 \subset P_2$.

\begin{example}
The following are relationships between arguments from Example~\ref{ex:args}, 
based on Definitions~\ref{def:pspecificity} and~\ref{def:MGEPref}:

\smallskip
\noindent
$\cala_1$ and $\cala_6$ are incomparable (blocking defeaters); \\
$\cala_6 \succ \cala_2$, and thus $\cala_6$ defeats $\cala_2$; \\
$\cala_6 \succ \cala_3$, and thus $\cala_6$ defeats $\cala_3$; \\
$\cala_6 \succ \cala_4$, and thus $\cala_6$ defeats $\cala_4$; \\
$\cala_5$ and $\cala_7$ are incomparable (blocking defeaters).
\smallbsq
\end{example}

A sequence of arguments called an \emph{argumentation line} thus arises from this attack relation,
where each argument defeats its predecessor. To avoid undesirable sequences, that may represent circular
or fallacious argumentation lines, in \DLP\ an \emph{argumentation line} is \emph{acceptable} if it
satisfies certain constraints (see~\cite{Garcia-Simari04-TPLP}).
A literal $L$ is \emph{warranted} if there exists a non-defeated argument \Arg\ supporting $L$.

Clearly, there can be more than one defeater for a particular argument $\<\A, L\>$. Therefore, many acceptable
argumentation lines could arise from $\<\A, L\>$, leading to a tree structure.
The tree is built from the set of all argumentation lines rooted in the initial argument.
In a dialectical tree, every node (except the root) represents a defeater of its parent, and leaves
correspond to undefeated arguments.
Each path from the root to a leaf corresponds to a different acceptable argumentation line.
A dialectical tree provides a structure for considering all the possible acceptable argumentation lines that
can be generated for deciding whether an argument is defeated. We call this tree \emph{dialectical} because
it represents an exhaustive dialectical\footnote{In the sense of providing reasons for and against a position.} analysis for the argument in its root.
For argument $\<\A, L\>$, we denote its dialectical tree with \Tree{\AL}.

Given a literal $L$ and an argument \AL, in order to decide whether or not a literal $L$ is warranted,
every node in the dialectical tree \Tree{\AL} is recursively marked as \Dnode\ (\emph{defeated}) or
\Unode\ (\emph{undefeated}), obtaining a marked dialectical tree \MTree{\AL} where:
\begin{itemize}
\item All leaves in \MTree{\AL} are marked as \Unode s, and

\item Let \Bq\ be an inner node of \MTree{\AL}.
Then, \Bq\ will be marked as \Unode\
      iff every child of \Bq\ is marked as  \Dnode.
Node \Bq\ will be marked as \Dnode\
    iff it has at least a child marked as \Unode.
\end{itemize}
Given argument \AL\ over $\PP$, if the root of \MTree{\AL} is marked \Unode, then
\MTree{\Ah} \emph{warrants} $L$ and that $L$ is \emph{warranted} from $\PP$.
(Warranted arguments correspond to those in the grounded extension of a Dung argumentation system \cite{dung95}.)

We can then extend the idea of a dialectical tree to a \textit{dialectical forest}.  For a given literal $L$, a dialectical forest $\calf(L)$ consists of the set of dialectical trees for all arguments for $L$.  We shall denote a marked dialectical forest, the set of all marked dialectical trees for arguments for $L$, as $\calfs(L)$.  Hence, for a literal $L$, we say it is \textit{warranted} if there is at least one argument for that literal in the dialectical forest $\calfs(L)$ that is labeled $\Unode$, \textit{not warranted} if there is at least one argument for literal $\neg L$ in the forest $\calfs(\neg L)$ that is labeled $\Unode$, and \textit{undecided} otherwise.

\section{The InCA Framework}
\label{sec:probarg}

Having defined our environmental and analytical models ($\PEM, \PAM$ respectively), we now define how the two relate,
which allows us to complete the definition of our InCA framework.

The key intuition here is that given a $\PAM$, every element of $\kitSink$ might only hold in certain worlds in the set
$\wldE$ -- that is, worlds specified by the environment model.  As formulas over the environmental atoms in set
$\GndE$ specify subsets of $\wldE$ (i.e., the worlds that satisfy them), we can use these formulas to identify the
conditions under which a component of $\kitSink$ \textit{can be} true.
Recall that we use the notation $\fEM$ to denote the set of
all possible formulas over $\GndE$.  Therefore, it makes sense to associate elements of $\kitSink$ with a formula from
$\fEM$.  In doing so, we can in turn compute the probabilities of subsets of $\kitSink$ using the information contained in
$\PEM$, which we shall describe shortly.  We first introduce the notion of \textit{annotation function},
which associates elements of $\kitSink$ with elements of $\fEM$.

We also note that, by using the annotation function (see Figure~\ref{fig:annoEx}), we may have certain statements that appear as both facts and presumptions
(likewise for strict and defeasible rules).  However, these constructs would have different annotations, and thus be
applicable in different worlds.  Suppose we added the following presumptions to our running example:
\begin{itemize}
\item[] $\p_3 = \evidOf(X,O) \; \defleftarrow$, and

\item[] $\p_4 = \motiv(X,X') \; \defleftarrow$.
\end{itemize}
Note that these presumptions are constructed using the same formulas as facts $\f_1,\f_2$.
Suppose we extend $\af$ as follows:
\begin{eqnarray*}
\af(\p_3)&=&\mwInOp(M,O) \wedge \mwRel(M,M') \\
&&\wedge \mwHint(M',X)\\
\af(\p_4)&=&\inLgConf(Y,X') \wedge \cooper(X,Y)
\end{eqnarray*}
So, for instance, unlike $\f_1$, $\p_3$ can potentially be true in any world of the form:
\[
\{\mwInOp(M,O), \mwRel(M,M'),\mwHint(M',X)\}
\]
while $\f_1$ cannot be considered in any those worlds.

With the annotation function, we now have all the components to formally define an InCA framework.

\begin{definition}[InCA Framework]
Given environmental model $\PEM$, analytical model $\PAM$, and annotation function $\af$, $\cali = (\PEM,\PAM,\af)$ is an \textbf{InCA framework}.
\end{definition}

\begin{figure}
\centering
\fbox{
\parbox{0.96\columnwidth}{
\begin{tabular}{ll}
$\af(\f_1)=$ & $\origIP(\worm,\baja) \vee $\\
						 & $\big(\mwInOp(\worm,o) \wedge$ \\
             & $\big( \mwHint(\worm,\baja) \vee$\\
						 & $(\compiledLan(\worm,c) \wedge$ \\
             & $\natLang(\baja,c))\big)\big)$\\
$\af(\f_2)=$ & $\inLgConf(\baja,\kras)$\\
$\af(\r_1)=$ & $\Tr$\\
$\af(\r_2)=$ & $\Tr$\\
$\af(\p_1)=$& $\numTtMse(\baja, 2) \vee\govCybCap(\baja)$\\
$\af(\p_2)=$ & $\mwInOp(\worm,o') \wedge$\\
&$\infSys(\kras,\worm)$\\
$\af(\p_3)=$ & $\cybCapAge(\baja, 5)$\\
$\af(\d_1)=$ & $\Tr$\\
$\af(\d_2)=$ & $\Tr$\\
$\af(\d_3)=$ & $\Tr$\\
$\af(\d_4)=$ & $\Tr$\\
$\af(\d_5)=$ & $\Tr$
\end{tabular}
}}
\caption{Example annotation function.}
\label{fig:annoEx}
\end{figure}

Given the setup described above, we consider a \textit{world-based} approach --
the defeat relationship among arguments will depend on the current state of the world (based on the EM).
Hence, we now define the status of an argument with respect to a given world.

\begin{definition}[Validity]
Given InCA framework\\ $\cali = (\PEM,\PAM,\af)$, argument $\< \A, L\>$ is valid w.r.t.\ world
$w \in \wldE$ iff $\forall c \in \A, w \models \af(c)$.
\end{definition}

In other words, an argument is valid with respect to $w$ if the rules, facts, and presumptions in that argument are present 
in $w$ -- the argument can then be built from information that is available in that world. In this paper, we extend the 
notion of validity to argumentation lines, dialectical trees, and dialectical forests in the expected way (an argumentation 
line is valid w.r.t.\ $w$ iff all arguments that comprise that line are valid w.r.t.\ $w$).

\begin{example}
\label{anoEx}
Consider worlds $w_1,\ldots,w_8$ from Example~\ref{linPrgEx} along with the argument $\<\A_5,\isCap(\baja,\worm)\>$ from Example~\ref{ex:args}.  This argument is valid in worlds $w_1$--$w_4$, $w_6$, and $w_7$.
\smallbsq
\end{example}

We now extend the idea of a dialectical tree w.r.t.\ worlds -- so, for a given world $w \in \wldE$, the
dialectical (resp., marked dialectical) tree induced by $w$ is denoted by $\calt_w{\AL}$ (resp., $\calts_w{\AL}$).
We require that all
arguments and defeaters in these trees to be valid with respect to $w$.  Likewise, we extend the notion of dialectical
forests in the same manner (denoted with $\calf_w(L)$ and $\calfs_w(L)$, respectively).  
Based on these concepts\, we introduce the notion of \textit{warranting scenario}.

\begin{definition}[Warranting Scenario]
\label{def:warrantScenario}
Let $\cali =$ 
$(\PEM,$ $\PAM,$ $\af)$ be an InCA framework and
$L$ be a ground literal over $\GndA$;
a world $w \in \wldE$ is said to be a {\em warranting scenario} for $L$
(denoted $w \war L$)
iff there is a dialectical forest $\calfs_w(L)$ in which $L$ is warranted and $\calfs_w(L)$ is valid w.r.t $w$.
\end{definition}

\begin{example}
Following from Example~\ref{anoEx}, argument
$\<\A_5,\isCap(\baja,\worm)\>$  is warranted in worlds
$w_3$, $w_6$, and $w_7$.
\smallbsq
\end{example}

Hence, the set of worlds in the EM where a literal $L$ in the AM \textit{must} be true is exactly the set of
warranting scenarios -- these are the ``necessary'' worlds, denoted:
\[
\must(L) = \{w \in \wldE \; | \; (w \war L).
\}
\]
Now, the set of worlds in the EM where AM literal $L$ \textit{can} be true is the following -- these are the
``possible'' worlds, denoted:
\[
\can(L) = \{ w \in \wldE \; | \; w \nwar \neg L \}.
\]
The following example illustrates these concepts.

\begin{example}
Following from Example~\ref{anoEx}: \\

\noindent
$\must(\isCap(\baja,\worm)) = \{w_3,w_6,w_7\}$ and \\

\noindent
$\can(\isCap(\baja,\worm)) = \{w_1,w_2,w_3,w_4,w_6,w_7\}$.

\smallbsq
\end{example}

Hence, for a given InCA framework $\cali$, if we are given a probability distribution $\Pr$ over the worlds in the EM,
then we can compute an upper and lower bound on the probability of literal $L$ (denoted $\Pdist$) as follows:

\[
\ell_{L,\Pr,\cali}= \sum_{w \in \must(L)}\Pr(w),
\]
\[
u_{L,\Pr,\cali} = \sum_{w \in \can(L)}\Pr(w),
\] and
\[
\ell_{L,\Pr,\cali} \leq \Pdist \leq  u_{L,\Pr,\cali}.
\]

Now let us consider the computation of probability bounds on a literal when we are given a knowledge base $\kE$
in the environmental model, which is specified in $\cali$, instead of a probability distribution over all worlds.
For a given world $w \in \wldE$, let $\for(w) = \big(\bigwedge_{a \in w}a\big) \wedge
\big(\bigwedge_{a \notin w}\neg a\big)$ --
that is, a formula that is satisfied only by world $w$.  Now we can determine the upper and lower bounds on the
probability of a literal w.r.t.\ $\kE$ (denoted $\Pkb$) as follows:
\[
\ell_{L,\cali} = \textsf{EP-LP-MIN}\left(\kE,\bigvee_{w \in \must(L)}\for(w)\right),
\]
\[
u_{L,\cali} = \textsf{EP-LP-MAX}\left(\kE,\bigvee_{w \in \can(L)}\for(w)\right),
\]
and
\[
\ell_{L,\cali} \leq \Pkb \leq  u_{L,\cali}.
\]
Hence, $\Pkb = \left(\ell_{L,\cali}+\frac{u_{L,\cali}-\ell_{L,\cali}}{2}\right) \pm
\frac{u_{L,\cali}-\ell_{L,\cali}}{2}$.

\begin{example}
Following from Example~\ref{anoEx}, argument $\<\A_5,\isCap(\baja,\worm)\>$, we can compute
$
\textbf{\textsf{P}}_{\isCap(\baja,\worm),\cali}
$
(where $\cali=(\Pi_{EM}',\Pi_{AM},\af))$.  Note that for the upper bound, the linear program we need to set up
is as in Example~\ref{linPrgEx}.  For the lower bound, the objective function changes to:
$
\min x_3+x_6+x_7
$.
From these linear constraints, we obtain:
$
\textbf{\textsf{P}}_{\isCap(\baja,\worm),\cali} = 0.75 \pm 0.25.
$
\smallbsq
\end{example}

\section{Attribution Queries}
\label{sec:attrib-queries}

We now have the necessary elements required to formally define the kind of queries that correspond to
the attribution problems studied in this paper.

\begin{definition}
\label{def:attribQuery}
Let $\cali = (\PEM,\PAM,\af)$ be an InCA framework,
$\S \subseteq \actors$ (the set of ``suspects''), $\O \in \ops$ (the ``operation''),
and $\E \subseteq \GndE$ (the ``evidence'').
An actor $\act \in \S$ is said to be a {\em most probable suspect}
iff there does not exist $\act' \in \S$ such that $\Pkbpco > \Pkbco$ where $\cali' = (\PEM\cup \Pi_{\E},\PAM,\af')$
with $\Pi_{\E}$ defined as $\bigcup_{c \in \E} \{c: 1 \pm 0\}$.
\end{definition}

Given the above definition, we refer to $Q = (\cali,\S,\O,\E)$ as an {\em attribution query},
and $\act$ as an {\em answer} to $Q$.  We note that in the above definition, the items of evidence are added to the
environmental model with a probability of $1$.  While in general this may be the case, there are often instances in analysis
of a cyber-operation where the evidence may be true with some degree of uncertainty.  Allowing for probabilistic evidence is a
simple extension to Definition~\ref{def:attribQuery} that does not cause any changes to the results of this paper.

To understand how uncertain evidence can be present in a cyber-security scenario, consider the following.  In Symantec's initial analysis of the Stuxnet worm, they found the routine designed to attack the S7-417 logic controller was incomplete, and hence
would not function~\cite{symantecStuxnet}.  However, industrial control system expert Ralph Langner claimed that the
incomplete code would run provided a missing data block is generated, which he thought was possible~\cite{langnerBlog}.
In this case, though the code was incomplete, there was clearly uncertainty regarding its usability.
This situation provides a real-world example of the need to compare arguments -- in this case, in the worlds where both
arguments are valid, Langner's argument would likely defeat Symantec's by generalized specificity (the outcome, of course,
will depend on the exact formalization of the two).  Note that Langner was later vindicated by the discovery of an older sample, Stuxnet 0.5, which generated the data block.\footnote{http://www.symantec.com/connect/blogs/stuxnet-05-disrupting-uranium-processing-natanz}

InCA also allows
for a variety of relevant scenarios to the attribution problem.  For instance, we can easily allow for the modeling of
non-state actors by extending the available constants -- for example, traditional groups such as Hezbollah, which has
previously wielded its cyber-warfare capabilities in operations against Israel~\cite{shakBook13}.
Likewise, the InCA can also be used to model cooperation among different actors in performing an attack,
including the relationship  between non-state actors and nation-states, such as the potential connection between Iran and
militants stealing UAV feeds in Iraq, or the much-hypothesized relationship between hacktivist youth groups and the Russian
government~\cite{shakBook13}.  Another aspect that can be modeled is deception where, for instance, an actor may leave false
clues in a piece of malware to lead an analyst to believe a third party conducted the operation.  Such a deception scenario can
be easily created by adding additional rules in the AM that allow for the creation of such counter-arguments.  Another type of
deception that could occur include attacks being launched from a system not in the responsible party's area, but under their
control (e.g., see \cite{shadow10}).  Again, modeling who controls a given system can be easily accomplished in our framework,
and doing so would simply entail extending an argumentation line.
Further, campaigns of cyber-operations can also be modeled, as well as relationships among malware and/or attacks (as
detailed in~\cite{mandiant}).  

As with all of these abilities, InCA provides the analyst the means to model a complex situation
in cyber-warfare but saves him from carrying out the reasoning associated with such a situation.  Additionally,
InCA results are constructive, so an analyst can ``trace-back'' results to better understand how
the system arrived at a given conclusion.

\section{Conclusion}
\label{sec:conc}

In this paper we introduced InCA, a new framework that allows the modeling of various cyber-warfare/cyber-security scenarios in order to help answer the attribution question by means of a combination of probabilistic modeling and argumentative reasoning.  This is the first framework, to our knowledge, that addresses the attribution problem while allowing for multiple pieces of evidence from different sources, including traditional (non-cyber) forms of intelligence such as human intelligence.  Further, our framework is the first to extend Defeasible Logic Programming with probabilistic information. Currently, we are implementing InCA and the associated algorithms and heuristics to answer these queries.  We also feel that there are some key areas to explore relating to this framework, in particular:
\begin{itemize}
\item Automatically learning the EM and AM from data.
\item Conducting attribution decisions in near real time.
\item Identifying additional evidence that must be collected in order to improve a given attribution query.
\item Improving scalability of InCA to handle large datasets.
\end{itemize}
Future work will be carried out in these directions, focusing on the use of both real and synthetic datasets for empirical
evaluations.

\section*{Acknowledgments}

This work was supported by UK EPSRC grant EP/J008346/1 -- ``PrOQAW'',
ERC grant 246858 -- ``DIADEM'', by NSF grant \#1117761, by the Army
Research Office under the Science of Security Lablet grant (SoSL) and
project 2GDATXR042, and DARPA project R.0004972.001.  

The opinions in this paper are those
of the authors and do not necessarily reflect the opinions of the
funders, the U.S. Military Academy, or the U.S. Army.

\bibliographystyle{IEEEtran}
\bibliography{ProbPreDeLP}
\end{document}